\documentstyle[11pt]{article}
\begin{document}
\begin{center}
{\Large \bf On Issues in Swiss Cheese Compactifications}\footnote{Invited Review for MPLA based on talks given at Cornell, Caltech, UCLA and UC Berkeley}
\vskip 0.1in { Aalok Misra\footnote{e-mail: aalokfph@iitr.ernet.in}\\
Department of Physics, Indian Institute of Technology,
Roorkee - 247 667, Uttaranchal, India}
\end{center}
\begin{abstract}
We give a brief review of our previous works: \cite{SwissCheeseissues,axionicswisscheese}. We discuss two sets of issues. The first has to do with  the possibility of getting a non-supersymmetric dS minimum without the addition of $\overline{D3}$-branes as in KKLT, and axionic slow-roll inflation, in type II flux compactifications. The second has to do with
the ``Inverse Problem" \cite{VafaInverse} and ``Fake Superpotentials" \cite{Ceresole+Dall'agata} for extremal (non)supersymmetric black holes in
type II compactifications. We use (orientifold of) a ``Swiss Cheese" Calabi-Yau \cite{SwissCheese} expressed as a degree-18 hypersurface in ${\bf WCP}^4[1,1,1,6,9]$ in the ``large-volume-scenario" limit \cite{Balaetal2} for the former.
\end{abstract}

\section{Introduction}

Flux compactifications have been extensively studied from the point of view of moduli stabilization
(See \cite{Grana} and references therein). Though, generically only the complex structure moduli get stabilized
by turning on fluxes and one needs to consider non-perturbative moduli stabilization for the K\"{a}hler
moduli\cite{KKLT}.
Another extremely important issue related to moduli stabilization is the problem of getting a non-supersymmetric de Sitter vacuum in string theory. The KKLT scenario which even though does precisely that, has the problem of addition of an uplift term to the the potential, corresponding to addition of $\overline{D3}$-branes, that can not be cast into an ${\cal N}=1$ SUGRA formalism. It would be interesting to be able to get a de Sitter vacuum without the addition of such $\overline{D3}$-branes. The Large Volume Scenarios' study initiated in \cite{Balaetal2} provides a hope for the same. Further,  there is a close connection between flux vacua and black-hole attractors. It has been shown that extremal black holes exhibit an interesting phenomenon - the attractor mechanism \cite{attractor1}-\cite{attractor8}. In the same, the moduli are ``attracted" to some fixed values determined by the charges of the black hole, independent of the asymptotic values of the moduli. Supersymmetric black holes at the attractor point, correspond to
minimizing the central charge and the effective black hole potential, whereas nonsupersymmetric attractors \cite{nonsusybh1}, at the
attractor point, correspond to minimizing only the potential and not the central charge. The latter have recently been (re)discussed \cite{nonsusybh2}-\cite{nonsusybh8} in the literature.

In \cite{SwissCheeseissues}, we addressed the issues of the previous paragraph by exploring different perturbative and non-perturbative (in $\alpha^\prime$ and instanton contributions) aspects of (non)supersymmetric flux vacua and
black holes in the context of type II compactifications on (orientifold) of compact Calabi-Yau's of a
projective variety with multiple singular conifold loci in their moduli space. The compact Calabi-Yau we work with is of the ``Swiss cheese" type.

The embedding of inflation in string theory has been a field of recent interest because of several attempts to construct inflationary models in the context of string theory to reproduce CMB and WMAP observations \cite{kallosh1,wmap,KKLMMT}. These Inflationary models are also supposed to be good candidates for ``testing"  string theory \cite{kallosh1}. Initially, the idea of inflation was introduced to explain some cosmological problems like horizon problem, homogeneity problem, monopole problem etc.\cite{FirstInflation,cosmoproblem,linde}. Some ``slow roll" conditions were defined (with ``$\epsilon$" and ``$\eta$" parameters) as sufficient conditions for inflation to take place for a given potential. In string theory it was a big puzzle to construct inflationary models due to the problem of stability of compactification of internal manifold, which is required for getting a potential which could drive the inflation and it was possible to {\it rethink} about the same only after the volume modulus (as well as complex structure and axion-dilaton) could be stabilized by introducing non-perturbative effects (resulting in a meta-stable dS also) \cite{KKLT}. Subsequently, several models have been constructed with different approaches such as ``brane inflation" (for example $ D3/\overline{D3}$ branes in a warped geometry, with the brane separation as the inflaton field, D3/D7 brane inflation model \cite{KKLMMT,kesav},\cite{braneinflation1}-\cite{braneinflation5}) and ``modular inflation" \cite{ book,alphacorrection,kahlerinflation}, but all these models were having the so called $\eta$- problem which was argued to be solved by fine tuning some parameters of these models. The models with multi scalar fields (inflatons) have also been proposed  to solve the $\eta$ problem \cite{Assisted}-\cite{Assisted3}. Meanwhile in the context of type IIB string compactifications, the idea of ``racetrack inflation" was proposed by adding an extra exponential term with the same  K\"{a}hler modulus but with a different weight in the expression for the superpotential (\cite{pillado2}). This was followed by ``Inflating in a better racetrack"  proposed by Pillado et al \cite{pillado1} considering two K\"{a}hler moduli in superpotential; it was also suggested that inflation may be easier to achieve if one considers more (than one) K\"{a}hler moduli. The potential needs to have a flat direction which provides a direction for the inflaton to inflate. For the multi-K\"{a}hler moduli, the idea of treating the ``smaller" K\"{a}hler modulus as inflaton field was also proposed \cite{kahlerinflation,kahler}. Recently, ``axionic inflation" in the context of type IIB compactifications shown by Grimm and Kallosh et al \cite{Grimm,AxionInflation}, seems to be of great interest for stringy inflationary scenarios \cite{AxionInflation}. In \cite{SwissCheeseissues}, we had shown the possibility of getting a dS vacuum {\it without} the addition of $\overline{D3}$-branes as in KKLT scenarios \cite{KKLT}, in type IIB ``Swiss Cheese" Calabi-Yau (See \cite{SwissCheese}) orientifold compactifications in the large volume limit. In \cite{axionicswisscheese}, developing further on this idea, we proposed the possibility of axionic inflation in the same model.

This review is planned as follows.
In section {\bf 2}, we show that by the inclusion of non-perturbative $\alpha^\prime$-corrections to
the K\"{a}hler potential that survive orientifolding and instanton contributions to the superpotential, one
can, analogous to \cite{Balaetal2}, get a large-volume non-supersymmetric dS vacuum {\it without the addition of $\overline{D3}$-branes}. We also include a discussion on one-loop and two-loop corrections to the K\"{a}hler potential. Further, we discuss the possibility of getting axionic inflation with the NS-NS axions providing the flat direction for slow roll inflation to proceed starting from a saddle point and proceeding towards the nearest dS minimum and show that it is possible to get the number of e-foldings to be 60. In section {\bf 3}, we explicitly solve the ``inverse problem" using the techniques of \cite{VafaInverse} and using the techniques of \cite{Ceresole+Dall'agata}, we show the existence of multiple superpotentials (including therefore ``fake superpotentials").
Section {\bf 4} has the conclusions.

\section{Non-supersymmetric dS minimum via Non-perturbative $\alpha^\prime$- and Instanton Corrections}

In this section, using the results of \cite{Grimm}, we show that after inclusion of non-perturbative
$\alpha^\prime$-corrections to the K\"{a}hler potential, in addition to the perturbative $\alpha^\prime$ corrections of \cite{BBHL},
as well as the non-perturbative instanton contributions to the superpotential, it may be possible to obtain a large volume
non-supersymmetric dS minimum (analogous to \cite{Balaetal2} for the non-supersymmetric AdS minimum) {\it without the addition of $\overline{D3}$-branes} - see
also \cite{westphal}.

Let us begin with a summary of the inclusion of perturbative $\alpha^\prime$-corrections to the K\"{a}hler potential in type IIB string theory
compactified on Calabi-Yau three-folds with NS-NS and RR fluxes turned on, as discussed in \cite{BBHL}. The
$(\alpha^\prime)^3-$ corrections contributing to the K\"{a}hler moduli space metric are contained in
\begin{equation}
\label{eq:nonpert1}
\int d^{10}x\sqrt{g}e^{-2\phi}\left(R + (\partial\phi)^2 + (\alpha^\prime)^3\frac{\zeta(3) J_0}{3.2^{11}}
+ (\alpha^\prime)^3(\bigtriangledown^2\phi) Q\right),
\end{equation}
where the ${\cal O}(R^4)$ $J_0$ and ${\cal O}(R^3)$ $Q$ are defined in \cite{BBHL}.
The perturbative world-sheet corrections to the hypermultiplet moduli space of Calabi-Yau three-fold compactifications of
type II theories are captured by the prepotential:
\begin{equation}
\label{eq:nonpert4}
F(X)=\frac{i}{3}\kappa_{abc}\frac{X^aX^bX^c}{X^0} + (X^0)^2
\xi,
\end{equation}
where the $(\alpha^\prime)^3$-corrections are contained in
$\xi\equiv-(\alpha^\prime)^3\frac{\chi(CY_3)\zeta(3)}{2}$, $\kappa_{abc}$ being the classical $CY_3$ intersection numbers.
Substituting (\ref{eq:nonpert4}) in $K=-ln\left[X^i{\bar F}_i + {\bar X}^iF_i\right]$ gives:
\begin{equation}
\label{eq:nonpert5}
K = - ln\left[-\frac{i}{6}(z^a - {\bar z}^a)(z^b - {\bar z}^b)(z^c - {\bar z}^c) + 4\xi\right].
\end{equation}
Truncation of ${\cal N}=2$ to ${\cal N}=1$, implying reduction of the quaternionic geometry to K\"{a}hler
geometry, corresponds to a K\"{a}hler metric which becomes manifest in K\"{a}hler coordinates: $T^a=\frac{1}{3}g^a+i\hat{V}^a,
\tau=l+ie^{-\phi_0}$, the hat denoting the Einstein's frame in which, e.g.,
$\hat{V}_a=e^{\phi_0}\left(\frac{1}{6}\kappa_{abc}v^bv^c\right)$, $v^a$ being the K\"{a}hler moduli, and
the K\"{a}hler potential is given by:
\begin{equation}
\label{eq:nonpert6}
K = - ln\left(-(\tau-{\bar\tau})\right) - 2 ln\left(\hat{{\cal V}} + \frac{1}{2}\xi e^{\frac{-3\phi_0}{2}}\right) -
ln\left(-i\int_{CY_3}\Omega\wedge{\bar\Omega}\right),
\end{equation}
substituting which into the ${\cal N}=1$ potential
$V = e^K\left(g^{i{\bar j}}D_iW{\bar D}_{\bar j}{\bar W} - 3 |W|^2\right)$ (one sums over all the moduli),
one gets:
\begin{eqnarray}
\label{eq:nonpert7}
& & V = e^K\Biggl[(G^{-1})^{\alpha{\bar\beta}}D_\alpha W D_{\bar\beta}{\bar W} + (G^{-1})^{\tau{\bar\tau}}D_\tau W
D_{\bar\tau}{\bar W}
- \frac{9\hat{\xi}\hat{{\cal V}}e^{-\phi_0}}{(\hat{\xi}-\hat{{\cal V}})(\hat{\xi} + 2\hat{{\cal V}})}(W{\bar D}_{\bar\tau}{\bar W}
+ {\bar W}D_\tau W)\nonumber\\
& & -3\hat{\xi}\frac{((\hat{\xi}) ^2 + 7\hat{\xi}\hat{{\cal V}} + (\hat{{\cal V}})^2)}{(\hat{\xi}-\hat{{\cal V}})
(\hat{\xi} + 2\hat{{\cal V}})^2}|W|^2\Biggr],
\end{eqnarray}
the hats being indicative of the Einstein frame - in our subsequent discussion, we will drop the hats for
notational convenience.
The structure of the $\alpha^\prime$-corrected potential shows that the no-scale structure is no longer preserved
due to explicit dependence of $V$ on $\hat{{\cal V}}$ and the $|W|^2$ term is not cancelled. In what follows,
we will be setting $2\pi\alpha^\prime=1$.

The type IIB Calabi-Yau orientifolds containing O3/O7-planes considered involve modding out by $(-)^{F_L}\Omega\sigma$ where
${\cal N}=1$ supersymmetry requires $\sigma$ to be a holomorphic and isometric involution:
$\sigma^*(J)=J,\ \sigma^*(\Omega)=-\Omega$. Writing the complexified K\"{a}hler form
$-B_2+iJ=t^A\omega=-b^a\omega_a+iv^\alpha\omega_\alpha$ where $(\omega_a,\omega_\alpha)$ form canonical
bases for ($H^2_-(CY_3,{\bf Z}), H^2_+(CY_3,{\bf Z})$), the $\pm$ subscript
indicative of being odd under $\sigma$, one sees that in the large volume limit of $CY_3/\sigma$,
contributions from large $t^\alpha=v^\alpha$ are exponentially suppressed, however the contributions
from $t^a=-B_a$ are not. Note that it is understood that $a$ indexes the {\bf real} subspace of {\bf real} dimensionality $h^{1,1}_-=2$; the  {\bf complexified} K\"{a}hler moduli correspond to $H^{1,1}(CY_3)$ with {\bf complex} dimensionality $h^{1,1}=2$ or equivalently real dimensionality equal to 4. So, even though $G^a=c^a-\tau b^a$ (for real $c^a$ and $b^a$ and complex $\tau$) is complex, the number of $G^a$'s is indexed by $a$ which runs over the real subspace $h^{1,1}_-(CY_3)$\footnote{To make the idea more explicit, the involution $\sigma$ under which the NS-NS two-form $B_2$ and the RR two-form $C_2$ are odd can be implemented as follows. Let $z_i, {\bar z}_i, i=1,2,3$ be the complex coordinates and the action of $\sigma$ be defined as: $z_1\leftrightarrow z_2, z_3\rightarrow z_3$; in terms of the $x_i$ figuring in the defining hypersurface in equation (1) on page 2, one could take $z_{1,2}=\frac{x_{1,2}^9}{x_5}$, etc. in the $x_5\neq0$ coordinate patch. One can construct the following bases $\omega^{(\pm)}$ of real two-forms of $H^2$ even/odd under the involution $\sigma$:
\begin{eqnarray}
\label{eq:bases}
& & \omega^{(-)}=\{\sum(dz^1\wedge d{\bar z}^{\bar 2} - dz^2\wedge d{\bar z}^{\bar 1}),
 i(dz^1\wedge d{\bar z}^{\bar 1} - dz^2\wedge d{\bar z}^{\bar 2})\}\equiv\{\omega^{(-)}_1,\omega^{(-)}_2\},\nonumber\\
& & \omega^{(+)}=\{\sum i(dz^1\wedge d{\bar z}^{\bar 2} + dz^2\wedge d{\bar z}^{\bar 1}),\sum i dz^1\wedge d{\bar z}^{\bar 1}\}\equiv\{\omega^{(+)}_1,\omega^{(+)}_2\}.
\end{eqnarray}
This implies that $h^{1,1}_+(CY_3)=h^{1,1}_-(CY_3)=2$ - the two add up to give 4 which is the {\bf real} dimensionality of $H^2(CY_3)$ for the given Swiss Cheese Calabi-Yau. As an example, let us write down $B_2\in{\bf R}$ as
\begin{eqnarray}
\label{eq:Bform}
B_2 & = & B_{1{\bar 2}}dz^1\wedge d{\bar z}^{\bar 2} + B_{2{\bar 3}}dz^2\wedge d{\bar z}^{\bar 3} + B_{3{\bar 1}}dz^3\wedge d{\bar z}^{\bar 1}  + B_{2{\bar 1}}dz^2\wedge d{\bar z}^{\bar 1} + B_{3{\bar 2}}dz^3\wedge d{\bar z}^{\bar 2} + B_{1{\bar 3}}dz^1\wedge d{\bar z}^{\bar 3}\nonumber\\
& & + B_{1{\bar 1}}dz^1\wedge d{\bar z}^{\bar 1}+ B_{2{\bar 2}}dz^2\wedge d{\bar z}^{\bar 2}+ B_{3{\bar 3}}dz^3\wedge d{\bar z}^{\bar 3}.
\end{eqnarray}
Now, using
(\ref{eq:bases}), one sees that by assuming $B_{1{\bar 2}}=B_{2{\bar 3}}=B_{3{\bar 1}}=b^1$, and $B_{1{\bar 1}}=-B_{2{\bar 2}}= i b^2, B_{3{\bar 3}}=0$, one can write $B_2=b^1\omega^{(-)}_1 + b^2\omega^{(-)}_2\equiv\sum_{a=1}^{h^{1,1}_-=2}b^a\omega^{(-)}_a$.}
As shown in \cite{Grimm}, based on the $R^4$-correction to the $D=10$ type IIB supergravity action
\cite{Green+Gutperle}
and the modular completion of ${\cal N}=2$ quaternionic geometry by summation over all $SL(2,{\bf Z})$ images
of world sheet corrections as discussed in \cite{Llanaetal},
the non-perturbative large-volume  $\alpha^\prime$-corrections that
survive the process of orientifolding of type IIB theories (to yield ${\cal N}=1$) to the
K\"{a}hler potential is given by (in the Einstein's frame):
\begin{eqnarray}
\label{eq:nonpert8}
& & K = - ln\left(-i(\tau-{\bar\tau})\right) -ln\left(-i\int_{CY_3}\Omega\wedge{\bar\Omega}\right)\nonumber\\
 & & - 2\ ln\Biggl[{\cal V} + \frac{\chi(CY_3)}{2}\sum_{m,n\in{\bf Z}^2/(0,0)}
\frac{({\bar\tau}-\tau)^{\frac{3}{2}}}{(2i)^{\frac{3}{2}}|m+n\tau|^3}\nonumber\\
& & - 4\sum_{\beta\in H_2^-(CY_3,{\bf Z})} n^0_\beta\sum_{m,n\in{\bf Z}^2/(0,0)}
\frac{({\bar\tau}-\tau)^{\frac{3}{2}}}{(2i)^{\frac{3}{2}}|m+n\tau|^3}cos\left((n+m\tau)k_a\frac{(G^a-{\bar G}^a)}{\tau - {\bar\tau}}
 - mk_aG^a\right)\Biggr]\nonumber\\
 & & +\frac{C^{KK\ (1)}_s(U_\alpha,{\bar U}_{\bar\alpha})\sqrt{\tau_s}}{{\cal V}\left(\sum_{(m,n)\in{\bf Z}^2/(0,0)}\frac{\frac{(\tau-{\bar\tau})}{2i}}{|m+n\tau|^2}\right)} + \frac{C^{KK\ (1)}_b(U_\alpha,{\bar U}_{\bar\alpha})\sqrt{\tau_b}}{{\cal V}\left(\sum_{(m,n)\in{\bf Z}^2/(0,0)}\frac{\frac{(\tau-{\bar\tau})}{2i}}{|m+n\tau|^2}\right)},
\end{eqnarray}
where $n^0_\beta$ are the genus-0 Gopakumar-Vafa invariants for the curve $\beta$ and
$k_a=\int_\beta\omega_a$,
and $G^a=c^a-\tau b^a$, the real RR two-form potential $C_2=C_a\omega^a$ and the real NS-NS two-form potential
$B_2=B_a\omega^a$. In (\ref{eq:nonpert8}), the first line and $-2\ ln({\cal V})$ are the tree-level contributions, the second (excluding the volume factor in the argument of the logarithm) and third lines are the perturbative and non-perturbative $\alpha^\prime$ corrections - $\{n^0_\beta\}$ are the genus-zero Gopakumar-Vafa invariants that count the number of genus-zero rational curves -  the fourth line is the 1-loop contribution; $\tau_s$ is the volume of the ``small" divisor and $\tau_b$ is the volume of the ``big" divisor. The loop-contributions arise from KK modes corresponding to closed string or 1-loop open-string exchange between $D3$- and $D7$-(or $O7$-planes)branes wrapped around the ``s" and ``b" divisors - note that the two divisors for
${\bf WCP}^4[1,1,1,6,9]$, do not intersect (See \cite{Curio+Spillner}) implying that there is no contribution from winding modes corresponding to strings winding non-contractible 1-cycles in the intersection locus corresponding to stacks of intersecting $D7$-branes wrapped around the ``s" and ``b" divisors. One sees from (\ref{eq:nonpert8}) that in the LVS limit, loop corrections are sub-dominant as compared to the perturbative and non-perturbative $\alpha^\prime$ corrections.

As pointed out in \cite{Grimm}, in (\ref{eq:nonpert8}),
one should probably sum over the orbits of the discrete
subgroup to which the symmetry group $SL(2,{\bf Z})$ reduces. Its more natural to write out the K\"{a}hler potential
and the superpotential in terms of the ${\cal N}=1$ coordinates $\tau, G^a$ and $T_\alpha$ where
$
T_\alpha = \frac{i}{2}e^{-\phi_0}\kappa_{\alpha\beta\gamma}v^\beta v^\gamma - (\tilde{\rho}_\alpha - \frac{1}{2}\kappa_{\alpha ab}c^a b^b)
-\frac{1}{2(\tau - {\bar\tau})}\kappa_{\alpha ab}G^a(G^b - {\bar G}^b),
$
where $\tilde{\rho}_\alpha$ being defined via $C_4$(the RR four-form potential)$=\tilde{\rho}_\alpha\tilde{\omega}_\alpha,
\tilde{\omega}_\alpha\in H^4_+(CY_3,{\bf Z})$.
Based on the action for the Euclidean $D3$-brane world volume (denoted by $\Sigma_4$) action \\
$iT_{D3}\int_{\Sigma_4} e^{-\phi}\sqrt{g-B_2+F}+T_{D3}\int_{\Sigma_4}e^C\wedge e^{-B_2+F}$, the nonperturbative
superpotential coming from a $D3$-brane wrapping a divisor $\Sigma\in H^4(CY_3/\sigma,{\bf Z})$ such that the unit
arithmetic genus condition of Witten \cite{Witten} is satisfied, will be proportional to
(See \cite{Grimm})
$e^{\frac{1}{2}\int_\Sigma e^{-\phi}(-B_2+iJ)^2-i\int_\Sigma(C_4-C_2\wedge B_2+\frac{1}{2}C_0B_2^2}
= e^{iT_\alpha\int_\Sigma\tilde{\omega}_\alpha}\equiv e^{in_\Sigma^\alpha T_\alpha},$
where $C_{0,2,4}$ are the RR potentials. Based on appropriate transformation properties of the superpotential under the shift symmetry and
the subgroup $\Gamma_S\subset SL(2,{\bf Z})$,
the non-perturbative instanton-corrected superpotential was shown in \cite{Grimm} to be:
\begin{equation}
\label{eq:nonpert9}
W = \int_{CY_3}G_3\wedge\Omega + \sum_{n^\alpha}\frac{\theta_{n^\alpha}(\tau,G)}{f(\eta(\tau))}e^{in^\alpha T_\alpha},
\end{equation}
where the holomorphic Jacobi theta function is given as $
\theta_{n^\alpha}(\tau,G)=\sum_{m_a}e^{\frac{i\tau m^2}{2}}e^{in^\alpha G^am_a};$
$m^2=C^{ab}m_am_b, C_{ab}=-\kappa_{\alpha^\prime ab}$, $\alpha=\alpha^\prime$
corresponding to that $T_\alpha=T_{\alpha^\prime}$ (for simplicity).

Now, for the ``Swiss cheese" \footnote{The term ``Swiss cheese" (See \cite{SwissCheese}) is used to denote those Calabi-Yau's whose volume can be written as: ${\cal V}=(\tau^B + \sum_{i\neq B} a_i\tau^S_i)^{\frac{3}{2}} - (\sum_{j\neq B}b_j\tau^S_j)^{\frac{3}{2}} - ...$, where $\tau^B$ is the volume of the big divisor and $\tau^S_i$ are the volumes of the $h^{1,1}-1$ (corresponding to the (1,$h^{1,1}-1$)-signature of the Hessian) small divisors. The big divisor governs the size of the Swiss cheese and the small divisors control the size of the holes of the same Swiss cheese.} Calabi-Yau three-fold obtained as a resolution of the degree-18 hypersurface in ${\bf WCP}^4[1,1,1,6,9]$:
\begin{equation}
\label{eq:hypersurface}
x_1^{18} + x_2^{18} + x_3^{18} + x_4^3 + x_5^2 - 18\psi \prod_{i=1}^5x_i - 3\phi x_1^6x_2^6x_3^6 = 0.
\end{equation}
Similar to the explanation given in \cite{Kachruetal}, it is understood that only two complex structure
moduli $\psi$ and $\phi$ are retained in (\ref{eq:hypersurface}) which are invariant under the group $G$ of footnote 3 of \cite{SwissCheeseissues}, setting the other invariant complex structure moduli appearing at a higher order (due to invariance under $G$)
 at their values at the origin. As shown in \cite{Denef+Douglas+Florea}, there are two divisors which
when uplifted to an elliptically-fibered Calabi-Yau, have a unit arithmetic genus (\cite{Witten}):
$\tau_1\equiv\partial_{t_1}{\cal V}=\frac{t^2_1}{2},\ \tau_2\equiv\partial_{t_2}{\cal V}=\frac{(t_1+6t_2)^2}{2}$.

As shown in \cite{SwissCheeseissues}, one gets the following potential:
\begin{eqnarray}
\label{eq:nonpert21}
& & V\sim\frac{{\cal Y}\sqrt{ln {\cal V}}}{{\cal V}^{2n^1+2}}e^{-2\phi}(n^1)^2\frac{\left(\sum_{m^a}e^{-\frac{m^2}{2g_s} + \frac{m_ab^a n^1}{g_s} + \frac{n^1\kappa_{1ab}b^ab^b}{2g_s}}\right)^2}{\left|f(\eta(\tau))\right|^2}
\nonumber\\
& & +\frac{ln {\cal V}}{{\cal V}^{n^1+2}}\left(\frac{\theta_{n^1}({\bar\tau},{\bar G})}{f(\eta({\bar\tau}))}
\right)e^{-in^1(-\tilde{\rho_1}+\frac{1}{2}\kappa_{1ab}
\frac{{\bar\tau}G^a-\tau{\bar G}^a}{({\bar\tau}-\tau)}\frac{(G^b-{\bar G}^b)}{({\bar\tau}-\tau)} -
\frac{1}{2}\kappa_{1ab}\frac{G^a(G^b-{\bar G}^b)}{(\tau-{\bar\tau})})}+c.c.\nonumber\\
& & +
\frac{|W|^2}{{\cal V}^3}\left(\frac{3k_2^2+k_1^2}{k_1^2-k_2^2}\right)
\frac{\left|\sum_c\sum_{n,m\in{\bf Z}^2/(0,0)}e^{-\frac{3\phi}{2}}A_{n,m,n_{k^c}}(\tau) sin(nk.b+mk.c)\right|^2}
{\sum_{c^\prime}\sum_{m^\prime,n^\prime\in{\bf Z}^2/(0,0)} e^{-\frac{3\phi}{2}}|n+m\tau|^3
|A_{n^\prime,m^\prime,n_{k^{c^{\prime}}}}(\tau)|^2 cos(n^\prime k.b+m^\prime k.c)}+\frac{\xi|W|^2}{{\cal V}^3}.
\nonumber\\
& &
\end{eqnarray}
On comparing (\ref{eq:nonpert21}) with the analysis of \cite{Balaetal2}, one sees that for generic values of
the moduli $\rho_\alpha, G^a, k^{1,2}$ and ${\cal O}(1)\ W_{c.s.}$, {\it and $n^1=1$}, analogous to \cite{Balaetal2}, the second term
dominates; the third term is a new term. However, as in KKLT scenarios (See \cite{KKLT}), $W_{c.s.}<<1$; we would henceforth assume that the fluxes and complex structure moduli have been so fine tuned/fixed that $W\sim W_{n.p.}$. Further, from studies related to study of axionic slow roll inflation in Swiss Cheese models \cite{SwissCheeseissues}, it becomes necessary to take $n^1>2$.  We assume that the fundamental-domain-valued $b^a$'s satisfy: $\frac{|b^a|}{\pi}<<1$\footnote{If one puts in appropriate powers of the Planck mass $M_p$, $\frac{|b^a|}{\pi}<<1$ is equivalent to $|b^a|<<M_p$, i.e., NS-NS axions are super sub-Planckian.}. This implies that the first term in (\ref{eq:nonpert21}) - $|\partial_{\rho^1}W_{np}|^2$ - a positive definite term and denoted henceforth by $V_I$, is the most dominant. Hence, if a minimum exists, it will be positive. To evaluate the extremum of the potential, one sees that:
\begin{eqnarray}
\label{eq:extrV-c_b}
& & \partial_{c^a}V_I\nonumber\\
& & \hskip -0.3in\sim- 4\frac{\sqrt{ln {\cal V}}}{{\cal V}^{2n^1+2}}\sum_{\beta\in H_2^-(CY_3,{\bf Z})} n^0_\beta\sum_{m,n\in{\bf Z}^2/(0,0)}mk^a
\frac{({\bar\tau}-\tau)^{\frac{3}{2}}}{(2i)^{\frac{3}{2}}|m+n\tau|^3} sin(n k.b + mk.c)\frac{\left(\sum_{m^a}e^{-\frac{m^2}{2g_s} + \frac{m_ab^a n^1}{g_s} + \frac{n^1\kappa_{1ab}b^ab^b}{2g_s}}\right)^2}{\left|f(\eta(\tau))\right|^2}=0\nonumber\\
& & \Leftrightarrow nk.b + mk.c = N\pi;\nonumber\\
& & \hskip-0.23in\partial_{b^a}V_I|_{nk.b + mk.c = N\pi}\sim\frac{{\cal V}\sqrt{ln {\cal V}}}{{\cal V}^{2n^1+1}}\frac{e^{-\frac{m^2}{2g_s} + \frac{m_{a^\prime}b^{a^\prime} n^1}{g_s} + \frac{n^1\kappa_{1a^\prime b^\prime}b^{a^\prime}b^{b^\prime}}{2g_s}}\sum_{m^a}e^{-\frac{m^2}{2g_s} + \frac{m_ab^a n^1}{g_s} + \frac{n^1\kappa_{1ab}b^ab^b}{2g_s}}}{\left|f(\eta(\tau))\right|^2}\left(\frac{n^1m^a}{g_s} + \frac{n^1\kappa_{1ab}b^b}{g_s}\right)=0.\nonumber\\
& & \end{eqnarray}
Now, given the ${\cal O}(1)$ triple-intersection numbers and super sub-Planckian NS-NS axions, we see that potential $V_I$ gets automatically extremized for $D1$-instanton numbers $m^a>>1$. Note that if the NS-NS axions get stabilized as per $\frac{n^1m^a}{g_s} + \frac{n^1\kappa_{1ab}b^b}{g_s}=0$, satisfying $\partial_{b^a}V=0$, this would imply that the NS-NS axions get stabilized at a rational number, and in particular, a value which is not a rational multiple of $\pi$, the same being in conflict with the requirement $nk.b + mk.c = N \pi$.
It turns out that the locus $nk.b + mk.c = N\pi$ for $|b^a|<<\pi$ and $|c^a|<<\pi$ corresponds to a flat saddle point with the NS-NS axions providing a flat direction - See \cite{axionicswisscheese}.

Analogous to \cite{Balaetal2}, for all directions in the moduli space with ${\cal O}(1)$ $W_{c.s.}$ and away from $D_iW_{cs}=D_\tau W=0=\partial_{c^a}V=\partial_{b^a}V=0$, the ${\cal O}(\frac{1}{{\cal V}^2})$ contribution
of $\sum_{\alpha,{\bar\beta}\in{c.s.}}(G^{-1})^{\alpha{\bar\beta}}D_\alpha W_{cs}{\bar D}_{\bar\beta}{\bar W}_{cs}$  dominates over (\ref{eq:nonpert21}),
ensuring that that there must exist a minimum, and given the positive definiteness of the potential $V_I$, this will be a dS minimum. There has been no need to add any $\overline{D3}$-branes as in KKLT to generate a dS vacuum. Also, interestingly, one can show that the condition $nk.b + mk.c = N \pi$ gurantees that the slow roll parameters ``$\epsilon$" and ``$\eta$" are much smaller than one for slow roll inflation beginning from the saddle point
and proceeding along an NS-NS axionic flat direction towards the nearest dS minimum (See \cite{axionicswisscheese}).

We now discuss the possibility of getting slow roll inflation along a flat direction provided by the NS-NS axions starting from a saddle point and proceeding to the nearest dS minimum. In what follows, we will assume that the volume moduli for the small and big divisors and the axion-dilaton modulus have been stabilized. All calculations henceforth will be in the axionic sector - $\partial_a$ will imply $\partial_{G^a}$ in the following. We need now to evaluate the slow-roll inflation parameters (in $M_p=1$ units)
$\epsilon\equiv\frac{{\cal G}^{ij}\partial_iV\partial_jV}{2V^2},\ \eta\equiv$ the most negative eigenvalue of the matrix $N^i_{\ j}\equiv\frac{{\cal G}^{ik}\left(\partial_k\partial_jV - \Gamma^l_{jk}\partial_lV\right)}{V}$.

The first derivative of the potential is given by:
\begin{equation}
\label{eq:dV}
\partial_aV|_{D_{c.s.}W=D_\tau W=0}=(\partial_a K)V+e^K\biggl[{\cal G}^{\rho_s{\bar\rho}_s}((\partial_a \partial_{\rho_s}W_{np} {\bar\partial}_{\bar\rho_s}){\bar W}_{np}+\partial_{\rho_s}W_{np}\partial_a{\bar\partial_{\bar\rho_s}}{\bar W}_{np})+\partial_a{\cal G}^{\rho_s{\bar\rho}_s}\partial_{\rho_s}W_{np}{\bar\partial}_{\bar\rho_s}{\bar W}_{np}\biggr].
\end{equation}
The most dominant terms in (\ref{eq:dV}) of ${\cal O}(\frac{\sqrt{ln {\cal V}}}{{\cal V}^{2n^s+1}})$ that potentially violate the requirement ``$\epsilon<<1$" are of the type:
(a)
e.g. $e^K(\partial_a{\cal G}^{\rho_s{\bar\rho}_s})(\partial_bW_{np}){\bar\partial}_{\bar c}{\bar W}_{np}$, is proportional to $\partial_a cos(nk.b + mk.c)$, which near the locus $nk.b + mk.c = N\pi$, vanishes;
(b)
e.g. $e^K {\cal G}^{\rho_s{\bar\rho}_s}\partial_a\partial_b W_{np}{\bar\partial}_{\bar c}{\bar W}_{np}$: the contribution to $\epsilon$ will be
$\frac{(n^s)^2e^{-\frac{2\alpha}{g_s}}{\cal V}}{\sum_{\beta\in H_2} (n^0_\beta)^2}$. We will choose ${\cal V}$ to be such that ${\cal V}\sim e^{\frac{2\alpha}{g_s}}$ (See \cite{LargeVcons}), implying that $\epsilon\sim\frac{(n^s)^2}{\sum_{\beta\in H_2}(n^0_\beta)^2}$. Now, it turns out that using the Castelnuovo's theory of moduli spaces that are fibrations of Jacobian of curves over the moduli space of their deformations, the genus-0 Gopakumar-Vafa integer invariants $n^0_\beta$'s for compact Calabi-Yau's of a projective variety in weighted complex projective spaces for appropriate degree of the holomorphic curve, can be as large as $10^{20}$ and even higher \cite{Klemm_GV} thereby guaranteeing that the said contribution to $\epsilon$ respects the slow roll inflation requirement. One can hence show from (\ref{eq:dV}) that near the locus $nk.b + mk.c = N\pi$, $\epsilon<<1$ is always satisfied.

To evaluate $N^a_{\  b}$ and the Hessian, one needs to evaluate the second derivatives of the potential and components of the affine connection. In this regard, one needs to evaluate, e.g.:
\begin{eqnarray}
\label{eq:ddV}
& & {\bar\partial}_{\bar d}\partial_aV=({\bar\partial}_{\bar d}\partial_aK)V+\partial_aK{\bar\partial}_{\bar d}V +e^K\biggr[{\bar\partial}_{\bar d}\partial_a{\cal G}^{\rho_s{\bar\rho_s}}\partial_{\rho_s}W_{np}{\bar\partial}_{\bar\rho_s}{\bar W}_{np}
+ \partial_a{\cal G}^{\rho_s{\bar\rho_s}}{\bar\partial}_{\bar d}\left(\partial_{\rho_s}W_{np}{\bar\partial}_{\bar\rho_s}{\bar W}_{np}\right)\nonumber\\
& & + {\bar\partial}_{\bar d}{\cal G}^{\rho_s{\bar\rho_s}}\partial_a\left(\partial_{\rho_s}W_{np}{\bar\partial}_{\bar\rho_s}{\bar W}_{np}\right) + {\cal G}^{\rho_s{\bar\rho_s}}\partial_a{\bar\partial}_{\bar d}\left(\partial_{\rho_s}W_{np}{\bar\partial}_{\bar\rho_s}{\bar W}_{np}\right)\biggr].
\end{eqnarray}
One can show that near the locus $nk.b + mk.c = N\pi$, the most dominant term (and hence the most dominant contribution to $\eta$) in (\ref{eq:ddV}) comes from
$e^K{\cal G}^{\rho_s{\bar\rho}_s}\partial_b\partial_{\rho_s}W_{np}{\bar\partial}_{\bar c}{\bar\partial_{\bar\rho_s}}{\bar W}_{np}$, proportional to:
$N^{\bar a}_{\ b}\ni \frac{{\cal V}n^sg_s\kappa}{\sum_{\beta\in H_2} (n^0_\beta)^2}.$ Having chosen ${\cal V}$ to be such that ${\cal V}\sim e^{\frac{2\alpha}{g_s}}$ , one gets $\eta\sim\frac{{\cal V}n^s\kappa}{ln {\cal V}\sum_{\beta\in H_2}(n^0_\beta)^2}$.
The large values of the genus-0 Gopakumar-Vafa invariants again nullifies this contribution to $\eta$.

Now, the affine connection components, in the LVS limit, are given by:
$\Gamma^a_{bc}={\cal G}^{a{\bar d}}\partial_b{\cal G}_{c{\bar d}}\sim\biggl[\left(\frac{{\bar\tau}}{{\bar\tau}-\tau}\right)\partial_{c^a}$\\
$+\left(\frac{1}{{\bar\tau}-\tau}\right)\partial_{b^a}\biggr]{\cal X}_1
\equiv {\cal O}({\cal V}^0),$
implying that
$N^{\bar a}_{\ b}\ni\frac{{\cal G}^{c{\bar a}}\Gamma^d_{cb}\partial_dV}{V}\sim\frac{{\cal V}
\sum_{m,n\in{\bf Z}^2/(0,0)}
\frac{({\bar\tau}-\tau)^{\frac{3}{2}}}{(2i)^{\frac{3}{2}}|m+n\tau|^3} sin(nk.b + mk.c)e^{-\frac{\alpha}{g_s}}}{\sum_{\beta\in H_2^-(CY_3,{\bf Z})} (n^0_\beta)^2}.$
We thus see that in the LVS limit and because of the large genus-0 Gopakumar-Vafa invariants, this contribution is nullified - note that near the locus $nk.b + mk.c = N\pi$, the contribution is further damped. Thus the ``$\eta$ problem" of \cite{KKLMMT} is solved.

One can show that near $nk.b+mk.c=N\pi$ and $b^a\sim-\frac{m^a}{\kappa}\sim\frac{N\pi}{nk^a}$, assuming that $\frac{nk.m}{\pi\kappa}\in{\bf Z}$, a direction in the NS-NS axionic space provides a nearly flat unstable direction for slow-roll inflation to commence.

The basis of axionic fields that would diagonalize the kinetic energy terms is given by:
\begin{equation}
\label{eq:diagbasis}
\left(\matrix{ \frac{{k_1}\,\left( {b^2}\,{k_1} - {b^1}\,{k_2} \right) \,
     {\sqrt{1 + \frac{{{k_2}}^2}{{{k_1}}^2}}}}{{{k_1}}^2 + {{k_2}}^2} \cr \frac{{k_1}\,
     \left( {c2}\,{k_1} - {c^1}\,{k_2} \right) \,
     {\sqrt{1 + \frac{{{k_2}}^2}{{{k_1}}^2}}}}{{{k_1}}^2 + {{k_2}}^2} \cr \frac{{k_2}\,
     \Omega_1\,
     \left( {b^1}\,\left( 1 + \left( -1 + A^2 \right) \,g_s^2 +
          {\sqrt{\cal S}} \right) \,{k_1} +
       A^2\,{b^2}\,g_s^2\,{k_2} + {b^2}\,
        \left( 1 - g_s^2 + {\sqrt{\cal S}} \right) \,
        {k_2} - 4\,A\,g_s^2\,\left( {c^1}\,{k_1} + {c2}\,{k_2} \right)  \right) }{4\,
     {\sqrt{2}}\,{\sqrt{\cal S}}\,
     \left( {{k_1}}^2 + {{k_2}}^2 \right) } \cr \frac{{k_2}\,
     \Omega_1\,
     \left( {b^1}\,\left( -1 - \left( -1 + A^2 \right) \,g_s^2 +
          {\sqrt{\cal S}} \right) \,{k_1} -
       A^2\,{b^2}\,g_s^2\,{k_2} + {b^2}\,
        \left( -1 + g_s^2 + {\sqrt{\cal S}} \right) \,
        {k_2} + 4\,A\,g_s^2\,\left( {c^1}\,{k_1} + {c2}\,{k_2} \right)  \right) }{4\,
     {\sqrt{2}}\,{\sqrt{\cal S}}\,
     \left( {{k_1}}^2 + {{k_2}}^2 \right) } \cr  }\right),
\end{equation}
where $\Omega_1\equiv {\sqrt{-\left( \frac{\left( -1 - \left( 1 + 14\,A^2 + A^4 \right) \,g_s^4 +
               {\sqrt{\cal S}} +
               \left( -1 + A^2 \right) \,g_s^2\,\left( -2 +
                  {\sqrt{\cal S}} \right)  \right) \,
             \left( {{k_1}}^2 + {{k_2}}^2 \right) }{A^2\,g_s^4\,{{k_2}}^2} \right) }}.$
This tells us that in the $g_s<<1$ limit, there are two NS-NS axionic basis fields in terms of which the axionic kinetic terms are diagonal -
${\cal B}^1\equiv\frac{\left( b^2k_1 - b^1k_2 \right)}
     {\sqrt{k_1^2 + k_2^2}}$, and
     ${\cal B}^2\equiv\frac{1}{2g_s\sqrt{2k_2^2(k_1^2+k_2^2)}}(b^1k_1 + b^2k_2)$.
By solving for $b^1$ and $b^2$ in terms of
${\cal B}^1$ and ${\cal B}^2$, and plugging into the mass term, one finds that the mass term for $B^2$ and not $B^1$, becomes proportional to $g_s^2(B^2)^2$ - given that the inflaton must be lighter than its non-inflatonic partner, one concludes that $\frac{1}{2g_s\sqrt{2k_2^2(k_1^2+k_2^2)}}(b^1k_1 + b^2k_2)$ must be identified with the inflaton. We need to consider a situation wherein one can not completely disregard $\frac{n^0_\beta}{\cal V}$ as compared to unity - this ratio could be smaller than unity but not negligible. This is because the eigenvalues and hence the eigenvectors of the Hessian are more sensitive to this ratio than the term $|{\cal X}|$ that one gets by assuming $\frac{n^0_\beta}{\cal V}<<1$ - in the latter case, one can show that one can not get a nearly flat unstable direction for slow roll to proceed.

To evaluate the number of e-foldings $N_e$, defining the inflaton ${\cal I}\sim b^2k_2 + b^1k_1$, one can show that (in $M_p=1$ units)
\[N_e=-\int_{{\rm in:\ Saddle\ Point}}^{{\rm fin:\ dS\ Minimum}}\frac{1}{\sqrt{\epsilon}}d{\cal I}\sim
  \frac{\sqrt{\sum_{\beta\in H_2}(n^0_\beta)^2}}{n^s\sqrt{{\cal V}}}\]
For appropriately high degree of the genus-0 holomorphic curve (usually 5 or more - See \cite{Klemm_GV}), one could choose $n^0_\beta$'s in such a way that $n^0_\beta\sim60n^s\sqrt{{\cal V}}$. This would yield the required 60 e-foldings.

\section{(Non-)Supersymmetric Black Hole Attractors}

We now switch gears and address two issues in the two subsections of this section, related to supersymmetric and non-supersymmetric black hole attractors. These are the ``inverse problem" and the existence of ``fake superpotentials".

\subsection{The ``Inverse Problem" for Extremal Black Holes}

 In this subsection,
using the techniques discussed in \cite{VafaInverse}, we explicitly solve the ``inverse problem" for extremal black holes in type II compactifications on (the mirror of) (\ref{eq:hypersurface}) - given a point in the moduli space, to find the charges $(p^I,q_I)$ that would satisfy $\partial_iV_{BH}=0$,
 $V_{BH}$ being the black-hole potential. In the next subsection, we address the issue of existence of ``fake superpotentials" in the same context.

 We will now summarize the ``inverse problem" as discussed in \cite{VafaInverse}.
 Consider $D=4, N=2$ supergravity coupled to $n_V$ vector multiplets in the absence of higher derivative
 terms. The black-hole potential can be written as \cite{nonsusybh1}:
 \begin{equation}
 \label{eq:BHinv1}
 V_{BH} = -\frac{1}{2}(q_I - {\cal N}_{IK} p^K)\left((Im {\cal N})^{-1}\right)^{IJ}(q_J - {\bar{\cal N}}p^L),
 \end{equation}
 where the $(n_V + 1)\times(n_V + 1)$ symmetric complex matrix, ${\cal N}_{IJ}$, the vector multiplet moduli space metric, is defined as:
 $ {\cal N}_{IJ} \equiv {\bar F}_{IJ} + \frac{2i Im(F_{IK}) X^K Im (F_{IL}) X^L}{Im(F_{MN}) X^M X^N},$
 $X^I,F_J$ being the symplectic sections and $F_{IJ}\equiv\partial_IF_J=\partial_JF_I$. The black-hole potential (\ref{eq:BHinv1}) can be rewritten (See \cite{VafaInverse}) as:
 \begin{equation}
 \label{eq:BHinv3}
 \tilde{V}_{BH} = \frac{1}{2}{\cal P}^I Im({\cal N}_{IJ}){\bar{\cal P}}^J - \frac{i}{2}{\cal P}^I(q_I - {\cal N}_{IJ}p^J)
 + \frac{i}{2}{\bar{\cal P}}^I(q_I - {\bar{\cal N}}_{IJ}p^J).
 \end{equation}
 The variation of (\ref{eq:BHinv3}) w.r.t. ${\cal P}^I$ gives:
 \begin{equation}
 \label{eq:BHinv4}
 {\cal P}^I=-i\left((Im {\cal N})^{-1})^{IJ}\right)(q_J - {\cal N}_{IJ}p^J),
 \end{equation}
 which when substituted back into (\ref{eq:BHinv3}), gives (\ref{eq:BHinv1}). From (\ref{eq:BHinv4}), one
 gets:
\begin{equation}
\label{eq:BHinv5}
p^I = Re({\cal P}^I),\ q_I = Re({\cal N}_{IJ}{\cal P}^J).
\end{equation}
Extremizing $\tilde{V}_{BH}$ gives:
\begin{equation}
\label{eq:BHinv6}
{\cal P}^I{\bar{\cal P}}^J\partial_i Im({\cal N}_{IJ}) + i({\cal P}^I\partial_i {\cal N}_{IJ} - {\bar{\cal P}}^J\partial_i{\bar{\cal N}}_{IJ})p^J = 0,
\end{equation}
which using (\ref{eq:BHinv5}) yields:
\begin{equation}
\label{eq:BHinv7}
\partial_i Im({\cal P}^I{\cal N}_{IJ}{\cal P}^J) = 0.
\end{equation}
Similar to what was done in section {\bf 3}, one uses the semi-classical approximation and
disregards the integrality of the electric and magnetic charges taking them to be large.

The inverse problem is not straight forward to define as all sets of charges $(p^I,q_I)$ which are related
to each other by an $Sp(2n_V + 2,{\bf Z})$-transformation, correspond to the same point in the moduli space. This is because the $V_{BH})$ (and $\partial_iV_{BH}$) is (are) symplectic invariants. Further, $\partial_iV_{BH}=0$ give $2n_V$ real equations in $2n_V+2$ real variables $(p^I,q_I)$. To fix these two
problems, one looks at critical values of $V_{BH}$ in a fixed gauge $W=w\in{\bf C}$. In other words,
\begin{equation}
\label{eq:BHinv8}
W=\int_M\Omega\wedge H = q_I X^I - p^I F_I = X^I(q_I - {\cal N}_{IJ}p^J) = w,
\end{equation}
which using (\ref{eq:BHinv5}), gives:
$
X^I Im({\cal N}_{IJ}){\bar{\cal P}}^J = w.$
Thus, the inverse problem boils down to solving:
\begin{eqnarray}
\label{eq:BHinv10}
& & p^I = Re({\cal P}^I),\ q_I=Re({\cal N}_{IJ}{\cal P}^J);\nonumber\\
& & \partial_i({\cal P}^I{\cal N}_{IJ}{\cal P}^J)=0,\ X^I{\cal N}_{IJ}{\bar{\cal P}}^J = iw.
\end{eqnarray}
One solves for ${\cal P}^I$s from the last two equations in (\ref{eq:BHinv10}) and substitutes the result
into the first two equations of (\ref{eq:BHinv10}).

We will now solve the last two equations of (\ref{eq:BHinv10}) for (\ref{eq:hypersurface}). As an example,
we work with points in the moduli space close to one of the two conifold loci: $\phi^3=1$. We need to work out the matrix $F_{IJ}$ so that one can work out the matrix ${\cal N}_{IJ}$. As shown in \cite{SwissCheeseissues} and using its notations, near $x=0$, one gets the following form of $F_{IJ}$:
\begin{equation}
\label{eq:BHinv12}
F_{IJ}=\left(\begin{array}{ccc} \frac{B_{01}}{C_3} + \frac{C_0}{C_3} & \frac{C_1}{C_3} & \frac{C_2}{C_3}\\
\frac{C_1}{C_3} & \frac{C_1}{C_4} & \frac{C_2}{C_4}\\
\frac{C_2}{C_3} & \frac{C_2}{C_4} & \frac{C_2}{C_5}
\end{array}\right).
\end{equation}
Using (\ref{eq:BHinv12}), one can evaluate $X^I Im(F_{IJ}) X^J$ - See \cite{SwissCheeseissues}. One hence obtains:
\begin{eqnarray}
\label{eq:BHinv15}
& & {\cal N}=\nonumber\\
& & \hskip -2cm\left(\begin{array}{ccc}
a_{00}+b^{(1)}_{00}x + b^{(2)}_{00} x ln x + c_{00}(\rho-\rho_0) &
a_{01}+b^{(1)}_{01}x + b^{(2)}_{01} x ln x + c_{01}(\rho-\rho_0) &
a_{02}+b^{(1)}_{02}x + b^{(2)}_{02} x ln x + c_{02}(\rho-\rho_0) \\
a_{01}+b^{(1)}_{01}x + b^{(2)}_{01} x ln x + c_{01}(\rho-\rho_0) &
a_{11}+b^{(1)}_{11}x + b^{(2)}_{11} x ln x + c_{11}(\rho-\rho_0) &
a_{12}+b^{(1)}_{12}x + b^{(2)}_{12} x ln x + c_{12}(\rho-\rho_0) \\
a_{02}+b^{(1)}_{02}x + b^{(2)}_{02} x ln x + c_{02}(\rho-\rho_0) &
a_{12}+b^{(1)}_{12}x + b^{(2)}_{12} x ln x + c_{12}(\rho-\rho_0) &
a_{22}+b^{(1)}_{22}x + b^{(2)}_{22} x ln x + c_{22}(\rho-\rho_0)
\end{array}\right).\nonumber\\
& &
\end{eqnarray}
The constants $a_{ij}, b^{(1),(2)}_{jk},c_{lm}$ are constrained by relations, e.g.,
$F_I={\cal N}_{IJ}X^J.$

So, substituting (\ref{eq:BHinv15}) into the last two equations of (\ref{eq:BHinv10}), one gets:
\begin{eqnarray}
\label{eq:BHinv17}
& & \partial_x({\cal P}^I{\cal N}_{IJ}{\cal P}^J)=0\Rightarrow ln x\left[({\cal P}^0)^2b^{(2)}_{00} + ({\cal P}^1)^2b^{(2)}_{11} + ({\cal P}^2)^2b^{(2)}_{22} + 2{\cal P}^0{\cal P}^1b^{(2)}_{01} + 2{\cal P}^0{\cal P}^2b^{(2)}_{02} + 2{\cal P}^1{\cal P}^2b^{(2)}_{12}\right]=0;\nonumber\\
& & \partial_{\rho-\rho_0}({\cal P}^I{\cal N}_{IJ}{\cal P}^J)=0\Rightarrow({\cal P}^0)^2c^{(2)}_{00} + ({\cal P}^1)^2c_{11} + ({\cal P}^2)^2c_{22} + 2{\cal P}^0{\cal P}^1c_{01} + 2{\cal P}^0{\cal P}^2c_{02} + 2{\cal P}^1{\cal P}^2c_{12}=0,
\end{eqnarray}
and ${\bar X}^I Im({\cal N}_{IJ}){\cal P}^J=-iw$ implies:
\begin{eqnarray}
\label{eq:BHinv18}
& & {\bar A}_I(a_{IJ}-{\bar a}_{IJ}){\cal P}^J+{\bar x}[{\bar B}_{I1}(a_{IJ} - {\bar a}_{IJ}){\cal P}^J
- {\bar b^{(1)}}_{IJ}{\bar A}_I{\cal P}^J] + x[b^{(1)}_{IJ}{\bar A}_I{\cal P}^J]
+ x ln x[{\bar A}_I b^{(2)}_{IJ}{\cal P}^J] + (\rho-\rho_0)[{\bar A}_I c_{IJ}{\cal P}^J] \nonumber\\
& & + ({\bar\rho} - {\bar\rho_0})[{\bar C}_I(a_{IJ} - {\bar a}_{IJ}){\cal P}^J - {\bar c}_{IJ}A_I{\cal P}^J]
+ {\bar x} ln{\bar x}[B_{I2}a_{IJ}{\cal P}^J]=-2{\bar w}\nonumber\\
& & {\rm or}\nonumber\\
& & \sum_{I=0}^2\Upsilon^I(x,{\bar x}, x ln x, {\bar x} ln {\bar x};\rho-\rho_0,{\bar\rho}-{\bar\rho_0}){\cal P}^I={\bar w}.
\end{eqnarray}
Using (\ref{eq:BHinv18}), we eliminate ${\cal P}^2$ from (\ref{eq:BHinv17}) to get:
\begin{equation}
\label{eq:BHinv19}
\alpha_i({\cal P}^0)^2 + \beta_i({\cal P}^1)^2 + \gamma_i{\cal P}^0{\cal P}^1 = \lambda_i,i=1,2.
\end{equation}
The equations (\ref{eq:BHinv19}) can be solved and yield four solutions which are:
\begin{eqnarray}
\label{eq:BHinv20}
& &
{\cal P}^0=\pm
\frac{1}{2\,{\sqrt{2}}\,\Biggl( {\alpha_2}\,{\lambda_1} - {\alpha_1}\,{\lambda_2} \Biggr) }\Biggl( {\gamma_2}\,{\lambda_1} - {\gamma_1}\,{\lambda_2} +
           \sqrt{Y} \Biggr)\sqrt{X},\  {{\cal P}^1}=\mp\frac{\sqrt{X}}{\sqrt{2}};\nonumber\\
& & {\cal P}^0=
\pm\frac{1}{2\,{\sqrt{2}}\,\Biggl( {\alpha_2}\,{\lambda_1} - {\alpha_1}\,{\lambda_2} \Biggr) }\Biggl( {\gamma_2}\,{\lambda_1} - {\gamma_1}\,{\lambda_2} -
           \sqrt{Y} \Biggr)\sqrt{X},\   {{\cal P}^1}=\mp\frac{\sqrt{X}}{\sqrt{2}}.
                         \end{eqnarray}
                         where $X, X_1$ and $Y$ are defined in \cite{SwissCheeseissues}.
 One can show that one does get ${\cal P}^I\sim X^I$ as one of the solutions - this corresponds to a supersymmetric
black hole, and the other solutions correspond to non-supersymmetric black holes.

\subsection{``Fake Superpotentials"}

In this subsection, using the results of \cite{Ceresole+Dall'agata}, we show the existence of ``fake superpotentials" corresponding to black-hole solutions
for type II compactification on (\ref{eq:hypersurface}).

As argued in \cite{Ceresole+Dall'agata}, dS-curved domain wall solutions in gauged supergravity and non-extremal black hole solutions in Maxwell-Einstein theory  have the same effective action. In the context of domain wall solutions, if there exists a ${\cal W}(z^i,{\bar z}^i)\in{\bf R}: V_{DW}(\equiv{\rm Domain\ Wall\ Potential})=-{\cal W}^2 + \frac{4}{3\gamma^2}g^{i{\bar j}}\partial_i{\cal W}\partial_{\bar j}{\cal W}$, $z^i$ being complex scalar fields, then the solution to the second-order equations for domain walls, can also be derived from the following first-order flow equations: $U^\prime=\pm e^U\gamma(r){\cal W};\ (z^i)^\prime = \mp e^U\frac{2}{\gamma^2}g^{i{\bar j}}\partial_{\bar j}{\cal W}$, where
$\gamma\equiv\sqrt{1 + \frac{e^{-2U}\Lambda}{{\cal W}^2}}$.

Now, spherically symmetric, charged, static and asymptotically flat black hole solutions of Einstein-Maxwell theory coupled to complex scalar fields have the form: $dz^2 = - e^{2U(r)} dt^2 + e^{-2U(r)}\biggl[\frac{c^4}{sinh^4(cr)} dr^2 $ $+ \frac{c^2}{sinh^2(cr)}(d\theta^2 + sin^2\theta d\phi^2)\biggr]$, where the non-extremality parameter
$c$ gets related to the positive cosmological constant $\Lambda>0$ for domain walls. For non-constant scalar fields, only for $c=0$ that corresponds to extremal black holes, one can write down first-order flow equations in terms of a ${\cal W}(z^i,{\bar z}^i)\in{\bf R}$: $U^\prime=\pm e^U{\cal W};\ (z^i)^\prime=\pm2e^U g^{i{\bar j}}\partial_{\bar j}{\cal W},$ and the potential $\tilde{V}_{BH}\equiv {\cal W}^2 + 4g^{i{\bar j}}\partial_i{\cal W}\partial_{\bar j}{\cal W}$ can be compared with the ${\cal N}=2$ supergravity black-hole potential $V_{BH}=|Z|^2+g^{i{\bar j}}D_iZD_{\bar j}{\bar Z}$ by identifying
${\cal W}\equiv|Z|$. For non-supersymmetric theories or supersymmetric theories where the black-hole constraint equation admits multiple solutions which may happen because several ${\cal W}$s may correspond to the same $\tilde{V}_{BH}$ of which only one choice of ${\cal W}$ would correspond to the true central charge, one hence talks about ``fake superpotential" or ``fake supersymmetry" - a ${\cal W}:\partial_i{\cal W}=0$ would correspond to a stable non-BPS black hole. Defining ${\cal V}\equiv e^{2U}V(z^i,{\bar z}^i),
{\cal\bf W}\equiv e^U{\cal W}(z^i,{\bar z}^i)$, one sees that ${\cal V}(x^A\equiv U,z^i,{\bar z}^i)=g^{AB}\partial_A{\bf W}(x)\partial_B{\bf W}(x)$, where $g_{UU}=1$ and $g_{Ui}=0$. This illustrates the fact that one gets the same potential ${\cal V}(x)$ for all vectors $\partial_A{\cal\bf W}$
with the same norm. In other words, ${\bf W}$ and $\tilde{\bf W}$ defined via: $\partial_A{\bf W}=R_A^{\ B}(z,{\bar z})\partial_B\tilde{\bf W}$ correspond to the same ${\cal V}$ provided:
$R^TgR=g$.

For ${\cal N}=2$ supergravity, the black hole potential $V_{BH}=Q^T{\cal M}Q$ where $Q=(p^\Lambda,q_\Lambda)$ is an $Sp(2n_v+2,{\bf Z})$-valued vector ($n_V$ being the number of vector multiplets) and ${\cal M}\in Sp(2n_V+2)$ is given by:
${\cal M}=\left(\begin{array}{cc}
A&B\\
C&D
\end{array}\right),$
where
$A\equiv Re {\cal N} (Im {\cal N})^{-1},\ B\equiv-Im {\cal N} - Re {\cal N} (Im {\cal N})^{-1} Re {\cal N} $ $C\equiv (Im {\cal N})^{-1} ,\ D = -A^T = - (Im {\cal N}^{-1})^T (Re {\cal N})^T.$
Define $M: {\cal M}={\cal I}M$ where $
M\equiv\left(\begin{array}{cc}
D&C\\
B&A
\end{array}\right),$
${\cal I}\equiv\left(\begin{array}{cc}
0&-{\bf 1}_{n_V+1}\\
{\bf 1}_{n_V+1}&0
\end{array}\right).$

The central charge $Z=e^{\frac{K}{2}}(q_\Lambda X^\Lambda - p^\Lambda F_\lambda)$, a symplectic invariant
is expressed as a symplectic dot product of $Q$ and covariantly holomorphic sections: ${\cal V}\equiv e^{\frac{K}{2}}(X^\Lambda,F_\Lambda)=(L^\Lambda,M_\Lambda) (M_\Lambda={\cal N}_{\Lambda\Sigma}L^\Sigma)$,
and hence can be written as
$Z = Q^T{\cal I}{\cal V} = L^\Lambda q_\Lambda - M_\lambda p^\Lambda.$
Now, the black-hole potential $V_{BH}=Q^T{\cal M}Q$ (being a symplectic invariant) is invariant under:
$Q\rightarrow SQ,\ S^T{\cal M}S={\cal M}.$
As $S$ is a symplectic matrix, $S^T{\cal I}={\cal I}S^{-1}$, this yields:
$[S,M]=0.$
In other words, if there exists a constant symplectic matrix $S:[S,M]=0$, then there exists a fake superpotential $Q^TS^T{\cal I}{\cal V}$ whose critical points, if they exist, describe non-supersymmetric
black holes.

We now construct an explicit form of $S$. For concreteness, we work at the point in the moduli space for
(\ref{eq:hypersurface}): $\phi^3=1$ and large $\psi$ near $x=0$ and $\rho=\rho_0$. Given the form of ${\cal N}_{IJ}$ in (\ref{eq:BHinv17}), one sees that:
$
M\equiv\left(\begin{array}{cc}
U & V \\
X & -U^T
\end{array}\right),
$
where $V^T=V,\ X^T=X$ and $U, V, X$ are $3\times 3$ matrices constructed from $Re {\cal N}$ and
$(Im {\cal N})^{-1}$. Writing
$
S=\left(\begin{array}{cc}
{\cal A} & {\cal B} \\
{\cal C} & {\cal D}
\end{array}\right),
$
(${\cal A}, {\cal B}, {\cal C}, {\cal D}$ are $3\times3$ matrices) and given that $S\in Sp(6)$, implying:
\begin{equation}
\label{eq:FakeW10}
\left(\begin{array}{cc}
{\cal A}^T & {\cal C}^T\\
{\cal B}^T & {\cal D}^T
\end{array}\right)\left(\begin{array}{cc}
0 & -{\bf 1}_3 \\
{\bf 1}_3 & 0
\end{array}\right)\left(\begin{array}{cc}
{\cal A} & {\cal B} \\
{\cal C} & {\cal D}
\end{array}\right)=\left(\begin{array}{cc}
0 & -{\bf 1}_3 \\
{\bf 1}_3 & 0
\end{array}\right).
\end{equation}
Now, $[S,M]=0$ implies:
\begin{equation}
\label{eq:FakeW12}
\left(\begin{array}{cc}
{\cal A} U + {\cal B} X & {\cal A} V - {\cal B} U^T \\
{\cal C} U + {\cal D} X & {\cal C} V - {\cal D} U^T
\end{array}\right) = \left(\begin{array}{cc}
U {\cal A} + V {\cal C} & U {\cal B} + V {\cal D} \\
X {\cal A} - U^T {\cal C} & X {\cal B} - U^T {\cal D}
\end{array}\right).
\end{equation}
The system of equations implied in (\ref{eq:FakeW10}) can be satisfied, e.g., by the following choice of ${\cal A}, {\cal B}, {\cal C}, {\cal D}$:
\begin{equation}
\label{eq:FakeW13}
{\cal B} = {\cal C}=0;\ {\cal D} = ({\cal A}^{-1})^T.
\end{equation}
To simplify matters further, let us assume that ${\cal A}\in O(3)$ implying that $({\cal A}^{-1})^T = {\cal A}$. Then (\ref{eq:FakeW12}) would imply:
\begin{equation}
\label{eq:FakeW14}
[{\cal A},V] = 0,\  [{\cal A},X] = 0,\ [{\cal A}^{-1},U] = 0,\ [{\cal A},U] = 0.
\end{equation}
For points near the conifold locus $\phi=\omega^{-1},\rho=\rho_0$, by dropping the moduli-dependent terms, one can show:
\begin{equation}
\label{eq:FakeW141}
\hskip -0.5cm  \left(Im {\cal N}^{-1}\right)_{0I}\left(Re {\cal N}\right)_{IK}=\left(Im {\cal N}^{-1}\right)_{0K}=\left(Im {\cal N}\right)_{0K} + \left(Re {\cal N}\right)_{0I}\left(Im {\cal N}^{-1}\right)_{IJ}
\left(Re{\cal N}\right)_{JK} = 0,\ K=1,2.
\end{equation}
This is equivalent to saying that
\begin{equation}
\label{eq:FakeW16}
S = \left(\begin{array}{cccccc}
1&0&0&0&0&0\\
0&-1&0&0&0&0\\
0&0&-1&0&0&0\\
0&0&0&1&0&0\\
0&0&0&0&-1&0\\
0&0&0&0&0&-1
\end{array}\right)
\end{equation}
is an allowed solution. We therefore see that the non-supersymmetric black-hole corresponding to the fake superpotential
$Q^TS^T{\cal I}{\cal V}$, $S$ being given by (\ref{eq:FakeW16}), corresponds to the change of sign of
two of the three electric and magetic charges as compared to a supersymmetric black hole. The symmetry
properties of the elements of ${\cal M}$ and hence $M$ may make it generically possible to find a constant
$S$ like the one in (\ref{eq:FakeW16}) for two-paramater Calabi-Yau compactifications.

\section{Conclusion}

In the large volume limit of ${\cal N}=1$ type IIB orientifold of  a Swiss-Cheese Calabi-Yau, we showed in \cite{SwissCheeseissues} that with the inclusion of
non-perturbative $\alpha^\prime$-corrections that survive the orientifolding alongwith the nonperturbative
contributions from instantons, it is possible to get a non-supersymmetric dS minimum
{\it without the inclusion of anti-D3 branes}. We generalized the idea in \cite{axionicswisscheese} of obtaining a dS minimum (using perturbative and non-perturbative corrections to the K\"{a}hler potential and instanton corrections to the superpotential) without the addition of $\overline{D3}$-branes by including the one- and two- loop corrections to the K\"{a}hler potential and showing that two-loop contributions are subdominant w.r.t. one-loop corrections and the one-loop corrections are sub-dominant w.r.t. the perturbative and non-perturbative $\alpha^\prime$ corrections in the {\it LVS} limits.
Assuming the NS-NS and RR axions $b^a, c^a$'s to lie in the fundamental-domain and to satisfy: $\frac{|b^a|}{\pi}<1,\ \frac{|c^a|}{\pi}<1$,  one gets a flat direction provided by the NS-NS axions for slow roll inflation to occur starting from a saddle point and proceeding to the nearest dS minimum.  The ``eta problem" gets solved at and away from the saddle point locus for some quantized values  of a linear combination of the NS-NS and RR axions; the slow-roll flat direction is provided by the NS-NS axions.
As regards supersymmetric and non-supersymmetric black-hole attractors
in ${\cal N}=2$ type II compactifications on the same Calabi-Yau three-fold, we explicitly solve the
``inverse problem" of determining the electric and magnetic charges of an extremal black hole given the
extremum values of the moduli. In the same context, we also show explicitly the existence of ``fake superpotentials"
as a consequence of non-unique superpotentials for the same black-hole potential corresponding to reversal
of signs of some of  the electric and magnetic charges.    A linear combination of the axions gets identified with the inflaton.

\end{document}